\documentclass{pasa}%

\usepackage{graphicx}

\title[Immersive virtual reality experiences for all-sky data sets]{Immersive virtual reality experiences for all-sky data}

\author[C.J.Fluke et al.]{C.J.~Fluke$^{1,2}$\thanks{cfluke@swin.edu.au} and D.G.~Barnes$^{3,4}$
\affil{$^1$Centre for Astrophysics \& Supercomputing, 
	Swinburne University of Technology, 
	Hawthorn, Victoria, 3122, Australia}
\affil{$^2$Advanced Visualisation Laboratory, Digital Research and Innovation Capability Platform,
	Swinburne University of Technology,
	Hawthorn, Victoria 3122, Australia}
\affil{$^3$Monash e-Research Centre, Monash University, Clayton, Victoria 3168, Australia}
\affil{$^4$Faculty of Information Technology, Monash University, Clayton, Victoria 3168, Australia}
}%

\jid{PASA}
\doi{xxx}
\jyear{\the\year}

\usepackage{aas_macros}
\usepackage{hyperref} 
\hypersetup{colorlinks,citecolor=blue,linkcolor=blue,urlcolor=blue}

\begin{document}

\begin{frontmatter}
\maketitle

\begin{abstract}
Spherical coordinate systems, which are ubiquitous in astronomy, cannot be shown without distortion on flat, two-dimensional surfaces.   This poses challenges for the two complementary phases of {\em visual exploration} -- making discoveries in data by looking for relationships, patterns or anomalies -- and {\em publication} -- where the results of an exploration are made available for scientific scrutiny or communication.   This is a long-standing problem, and many practical solutions have been developed.  Our {\tt allskyVR} approach provides a workflow for experimentation with commodity virtual reality head-mounted displays.  Using the free, open source {\sc s2plot} programming library, and the {\sc A-Frame} {\sc WebVR} browser-based framework, we provide a straightforward way to visualise all-sky catalogues on a user-centred, virtual celestial sphere.   The {\tt allskyVR} distribution contains both a quickstart option, complete with a gaze-based menu system, and a fully customisable mode for those who need more control of the immersive experience.  
The software is available for download from:
\url{https://github.com/cfluke/allskyVR}
\end{abstract}

\begin{keywords}
Visualisation -- virtual reality -- astronomical software -- catalogues
\end{keywords}
\end{frontmatter}

\section{INTRODUCTION }
\label{sec:intro}
To many ancient astronomers and skywatchers, the night sky was an enormous, material sphere surrounding 
the Earth.   Emerging from the natural philosophy of the pre-Socratics (c. 6th century BCE) was a description of 
stars attached to, and hence moving daily with this physical celestial orb.  Observing and explaining the motion of planets with respect to the fixed stars or identifying transient ``guest stars''  -- fundamental steps on the road to modern astronomy --  relied on methods to document and report celestial positions on a sphere.    

Today, a spherical coordinate pair, such as Right Ascension ($\alpha$) and Declination ($\delta$), is still the most convenient way to catalogue the locations of celestial objects.\footnote{We use $\alpha$ and $\delta$ to refer to {\em any} spherical coordinate pair.}  An additional celestial coordinate (distance, redshift or velocity), is encoded by mapping to a set of concentric spheres with differing radii.

Unfortunately, spherical coordinates provide a direct challenge when producing static plots, maps or charts to appear in flat, two-dimensional (2D) images -- the predominant method for analysing, intepreting, documenting and communicating scientific outcomes.  Without a spherical surface to print on and distribute, any 2D projection of the sky requires a compromise in accuracy between the scale, area, and azimuth of plotted positions -- all three properties cannot be presented simultaneously \citep{Farmer38}.   Instead, a decision must always be made (consciously or not) as to which subset of these properties are the most important, and which will be shown in a distorted fashion.    

This poses problems for the two complementary phases of {\em visual exploration} -- making discoveries in data by looking for relationships, patterns or anomalies -- and {\em publication} -- where the results of an exploration are made available for scientific scrutiny, education or public communication.  
\subsection{Visualising all-sky data}

The conventional approach to the problem of displaying, presenting or publishing data in 
spherical coordinates is to perform a mapping to a flat, 2D representation.
Usually developed for building better 2D maps of the (almost)  spherical Earth, a variety of 
projection techniques have made their way into astronomy.    

Computer-based plotting has 
vastly simplified the process of creating coordinate grids, or graticules, so that it is a 
straightforward task to implement different map projections for a particular data set.  \citet{Calabretta02} provide a comprehensive overview of map projections for astronomy, including forward and inverse transforms from celestial  to Cartesian coordinates.   

A number of browser-based tools for navigating all-sky data sets exist.  These include solutions that are: 
\begin{itemize}
\item Mainly intended for education and outreach, such as Google Sky\footnote{\url{https://www.google.com.au/sky/}} \citep{Connolly08} and WikiSky\footnote{\url{http://www.wikisky.org}}; 
\item Hybrid solutions merging a strong educational focus with direct links to the underlying data and publications, such as the WorldWide Telescope\footnote{\url{http://www.worldwidetelescope.org}} \citep{Goodman12,Fay16}; and \item Dedicated astronomical services, in particular Aladin Desktop\footnote{\url{http://aladin.u-strasbg.fr}} and Aladin Lite\footnote{\url{http://aladin.u-strasbg.fr/AladinLite/}} \citep{Boch14}.  
\end{itemize}

Using imagery from a variety of multi-wavelength surveys and observations, and offering control over features such as grid lines, constellation maps, and queries for an object of interest (e.g. by name, position, catalogue number, etc.) these solutions provide a powerful means to explore relationships between objects on the sky.  However, they suffer from map projection effects on large-scales, and are designed for viewing on flat, two-dimensional displays.

The Java-based {\sc topcat} \citep{Taylor05} package provides a comprehensive set of visualisation and analysis tools for catalogue data, including an interactive outside-looking-in all sky representation via the Spherical Polar Plot window.   {\sc topcat} supports a range of common astronomical data formats, and integration with Virtual Observatory services.

\citet{Kent17} provides a detailed discussion of the use of {\sc Blender}\footnote{\url{https://www.blender.org}}  and the Google Spatial Media module\footnote{\url{https://github.com/google/spatial-media}} to produce navigable spherical panoramas from astrophysical data, including FITS-format images, planetary terrain data and 3D catalogues.  The resulting videos can be viewed interactively via YouTube\footnote{E.g. \url{https://www.youtube.com/user/VisualizeAstronomy}} using a compatible browser.   A similar approach, using panoramic images generated from the {\sc splash} \citep{Price07} smoothed particle hydrodynamics code, was presented by \citet{Russell17}.

\subsection{Domes and head-mounted displays}

The astronomy education world has had a solution to the problem of spherical coordinate systems for some time: the planetarium dome.  Capable of dynamically 
representing $2 \pi$ steradians of the sky,  catalogues of objects can be shown at their correct location and correct angular separation without areal distortion in the coordinate system.  For aesthetic purposes, individual objects can be displayed with an exaggerated local scale, appearing much larger on the planetarium sky than we would ever see them unaided.

Despite continuous improvements in digital full-dome projection techniques, few professional astronomers spend their day making discoveries by projecting their data onto a dome \citep[see, for example, ][for early work]{Teuben01,Abbott04,Fluke06}.    The majority perform their day-to-day data exploration on desktop, notebook or tablet screens, accessing a much smaller solid angle.

The emergence of the consumer, virtual reality (VR) head-mounted display (HMD) presents a low-cost alternative to these large-scale  spaces.  
HMDs are ideal for providing an immersive, $4 \pi$ steradian, all-sky representation, where the viewer can look anywhere: forwards, backwards, up, down and side-to-side.   In essence, they provide a virtual, portable,  planetarium dome.

Two broad classes of commodity HMDs exist:
\begin{enumerate}
\item {\em Compute-based}: in the first generation, the head-set is connected to a computer via a cable, offering higher resolution, and real-time graphics.  This is often augmented with both basic orientation tracking via accelerometers and absolute  position tracking within a limited region.  Commercial options include the Oculus Rift\footnote{\url{https://www.oculus.com}}, HTC Vive\footnote{\url{https://www.vive.com}}, and Sony's PlayStation VR\footnote{\url{http://www.playstation.com/playstation-vr}}.  Increasingly, the computer, display system and an outward facing camera are combined into a single wearable, offering an experience where the digital world and the real world merge.  Often referred to as mixed reality (in contrast to the completely digital environment of VR), commercial and developer products are marketed by Microsoft (Hololens\footnote{\url{https://www.microsoft.com/hololens}}) and Google (the standalone Daydream\footnote{\url{https://vr.google.com/daydream/}}).
\item {\em Mobile-based}:  a smartphone is docked within a simple headset.  For experiences more complex than viewing a static image or short animation, it may be necessary to stream content to the phone via WiFi.  This places a limit on the frame-rates and level of interactivity that can be achieved.  Options here include Google Cardboard\footnote{\url{https://vr.google.com/cardboard/}} viewers and Samsung Gear VR\footnote{\url{http://www.samsung.com/global/galaxy/gear-vr/}}.
\end{enumerate}

\begin{figure*}
\includegraphics[width=0.99\linewidth,angle=0]{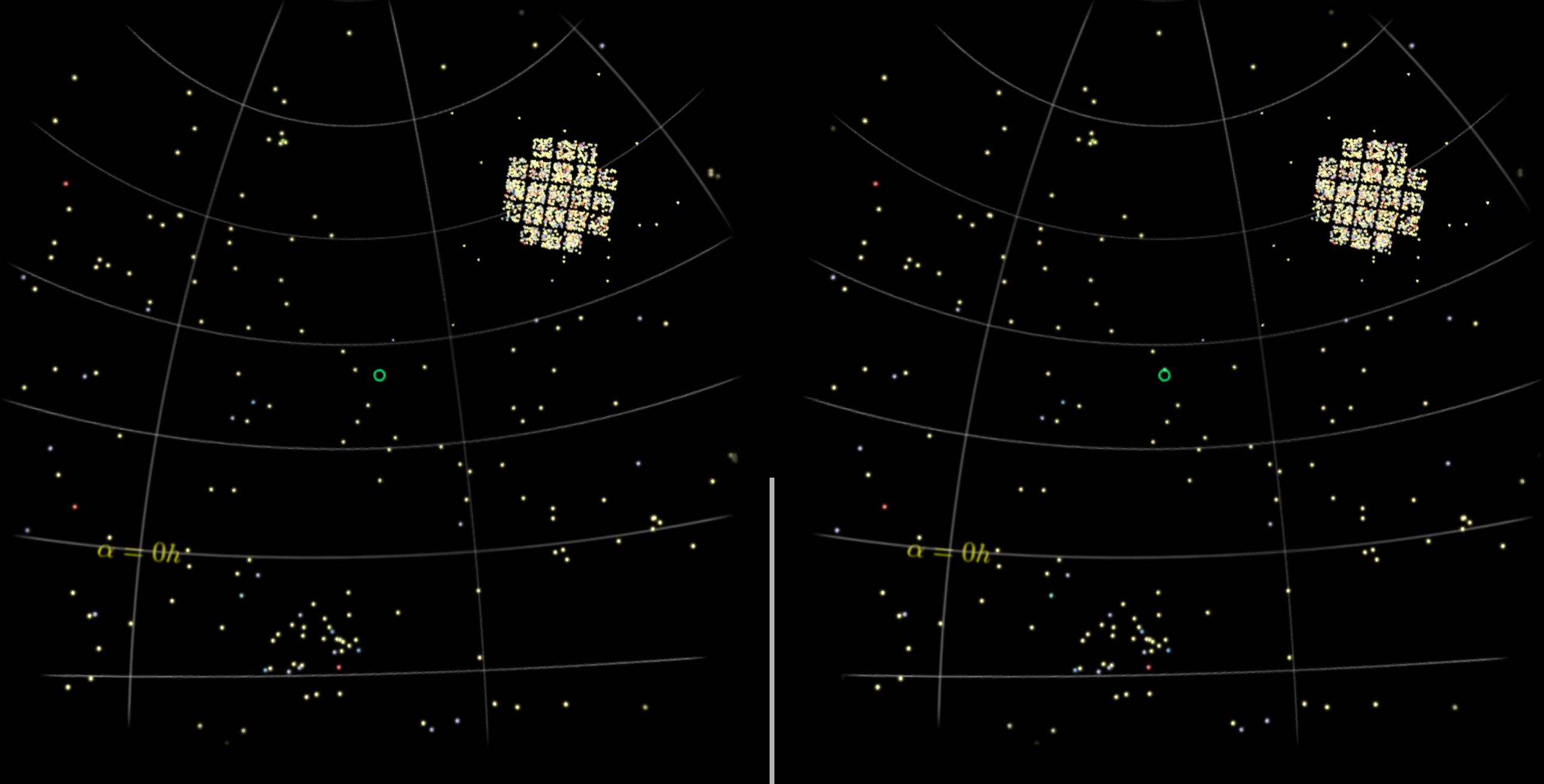}
\includegraphics[width=0.99\linewidth,angle=0]{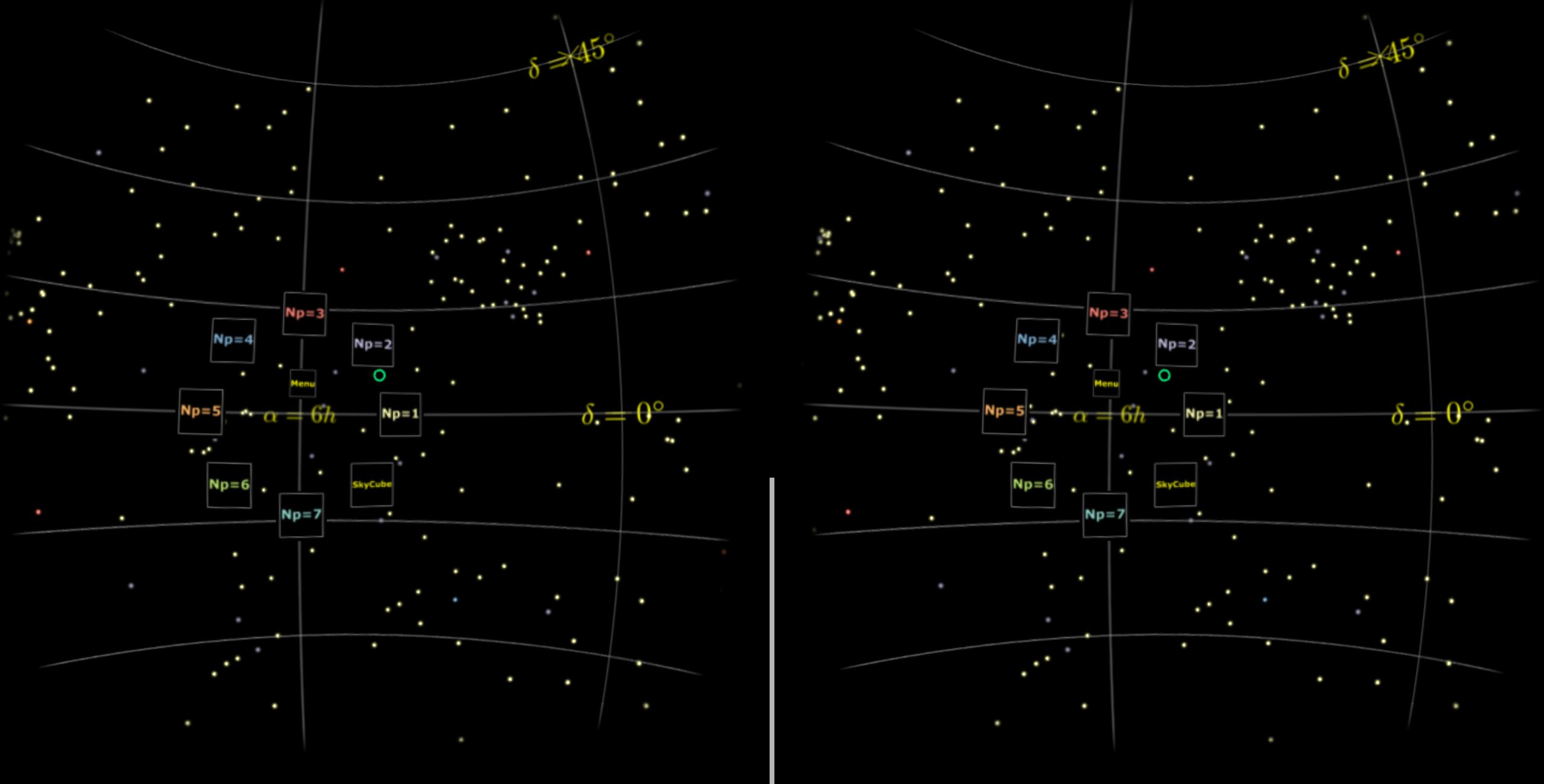}
\caption{Screenshots from an immersive, all-sky visualisation of confirmed exoplanetary systems 
using our {\sc s2plot} to {\sc A-Frame} export pathway.  The data set from the Kepler space mission is described in more detail in Appendix \ref{sct:example1}.  Features in the virtual reality environment include mapping of individual objects to {\sc A-Frame} entities (Section \ref{sct:prim}), a low-polygon count Sky Cube providing a reference grid (Section \ref{sct:SkyCube}), and a gaze-based menu system (Section \ref{sct:menu}). 
These two screenshots were captured from a Samsung Galaxy S7 Edge mobile device showing the left and right image views that form the immersive environment when viewed from a compatible head-mounted display. The expanded menu system is visible in the lower panel.  The vertical white line in the centre of each image is used to help with correct placement of the mobile device in a head-mounted display.  }
\label{fig:VRexo}
\end{figure*}

\subsection{Our solution}
In this Paper,  
we present a workflow for experimentation with commodity VR HMDs, targeted at all-sky catalogue data in spherical coordinates.  Our solution uses the free, open source {\sc s2plot} programming library\footnote{\url{http://astronomy.swin.edu.au/s2plot}} \citep[][and see Appendix \ref{app:depend}]{Barnes06} and the {\sc A-Frame} {\sc WebVR} framework initiated by MozillaVR\footnote{\url{https://mozvr.com}} (Section \ref{sct:aframe}).   

{\sc s2plot} offers a simple, but powerful, application programming interface (API) for 3D visualisation that hides access to {\sc OpenGL}\footnote{\url{https://www.opengl.org}} function calls.   The choice of display mode is a runtime decision depending on where the code is deployed.   {\sc s2plot} is written in C, and full functionality can be accessed most readily from C/C++ programs.  

{\sc A-Frame} visualisations are generated with {\sc html} statements, enriched by JavaScript functions, and can be viewed with a variety of commodity HMDs, including products from Oculus, HTC, Samsung and Google.

The {\tt allskyVR} system we present in this paper is a free, open source software solution for creating immersive, all-sky virtual reality experiences.  The distribution contains all of the relevant source code, scripts, and {\sc A-Frame} assets.  The software is available for download from:
\begin{center}
\url{https://github.com/cfluke/allskyVR}
\end{center}
with online documentation at:
\begin{center}
\url{https://http://allskyvr.readthedocs.io}
\end{center}

The VR environment supports orientation based navigation of the space, and a gaze-based menu system linked to a hierarchy of graphical entities.   The resulting assets can be transferred to a web server for online accessibility, most readily via a smartphone and Google Cardboard viewer.

\begin{figure*}
\begin{verbatim}
<head>
  <script src="https://aframe.io/releases/0.5.0/aframe.min.js"></script>
</head>
<body>
   <a-scene>
      <a-sky id="wrap" radius="100" material="color: #119"></a-sky>

      <a-entity id="Spheres-1" geometry="primitive:sphere; radius: 1.00; 
        segmentsWidth: 4; segmentsHeight: 4"
        material="wireframe: false; color: #ff0000" position="5.0 0.0 -10.0"></a-entity>
      <a-entity id="Spheres-2" geometry="primitive:sphere; radius: 1.00; 
        segmentsWidth: 4; segmentsHeight: 4"
        material="wireframe: false; color: #ff0000" position="-5.0  0.0 -10.0"></a-entity>
      <a-entity id="Spheres-3" geometry="primitive:sphere; radius: 1.00; 
        segmentsWidth: 4; segmentsHeight: 4"
        material="wireframe: false; color: #ff0000" position="0.0 5.0 -10.0"></a-entity>
      <a-entity id="Spheres-4" geometry="primitive:sphere; radius: 1.00; 
        segmentsWidth: 4; segmentsHeight: 4"
        material="wireframe: false; color: #ff0000" position="0.0 -5.0 -10.0"></a-entity>
   </a-scene>
</body>
\end{verbatim}
\caption{A simple {\sc A-Frame} scene. Four red spheres placed in front of the viewer.  The environment is surrounded by a dark blue spherical sky.   The polygon count for each of the sphere primitives is controlled by the {\tt segmentsWidth} and {\tt segmentsHeight}  parameters.  }
\label{fig:entities}
\end{figure*}

\section{Generating all-sky virtual reality environments}
\label{sct:aframe}
In this Section, we demonstrate how the {\tt allskyVR} distribution generates a virtual reality experience for viewing with a head-mounted display.  With a wide-variety of commodity HMDs available, we choose to focus on a straightforward approach to early adoption.

We use a viewing paradigm that maps the the positions of objects around the viewer onto a virtual celestial sphere, with the added constraint that the viewer's location is fixed to the sphere's origin.  This environment can be explored by looking in different directions, and is suitable for all HMDs that support head-orientation navigation.\footnote{For mobile-based HMDs, head-orientation navigation requires a smartphone equipped with an accelerometer.}  More complex, fully-navigable (e.g. with a handheld controller or through absolute position tracking) experiences are left to others to investigate.   However, we note that allowing the viewer to move away from the coordinate origin introduces new distortions -- the very problem we are trying to avoid.

In this work, an all-sky data set comprises: a spherical coordinate pair, $(\alpha, \delta)$, in decimal degrees; a radial coordinate ($r = 1$ for objects on the celestial sphere) in arbitrary units; an optional category index, which can be used to group objects with similar properties; and an optional per object scaling factor.    Colours are assigned to objects based on their category.

A ready to view astronomical example using our approach is included with the {\tt allskyVR} distribution.  Figure \ref{fig:VRexo} contains two screenshot from this example, incorporating all of the features that will be described in this Section.   The data set is  from the Kepler space mission, showing the locations of confirmed exoplanetary systems.   This data set is described in more detail in Appendix \ref{sct:example1}.

\subsection{{\tt allskyVR}}
{\tt AllskyVR} supports two modes of operation: {\em Quickstart} and {\em Customisable}.

Quickstart mode uses a subset of the {\tt allskyVR} distribution along with a set of assets we have generated in advance, thus reducing the barrier to experimentation and adoption.   Some level of customisation is still available, but only requiring direct modification of {\sc html} files or user-generated images for the menu system.  

A format conversion from input data to {\sc A-Frame} entities is performed, and a hierarchical framework and gaze-based menu system for selecting individual data categories is built.  The output is a collection of {\sc html} assets, images, and Javascript, along with a pre-generated coordinate grid.  This  mode can be used without installing {\sc s2plot}, and is best suited to catalogues with no more than a few thousand items.    

Customisable mode requires installation of the {\sc s2plot} programming library, third-party dependencies (see Appendix \ref{app:depend}), and the {\tt allskyVR} distribution.  Now, more complete control is possible over the appearance of graphical features such as category labels or choice of colour maps.   

Immediately following import of a data file into a custom {\sc s2plot} application, a keypress combination, {\tt <shift>-v}, initiates the workflow that creates the immersive experience.   The output comprises a complete set of assets:  {\sc html}  source, Javascript, and images.

The customisable mode provide functions to generate coordinate grids of constant right ascension and declination lines,  with configurable abels and other annotations.   Publication quality axis labels and annotations are generated using either the FreeType font library\footnote{\url{https://www.freetype.org}} or via \LaTeX\-style mathematical statements. 

For both modes, the resultant assets then need to be moved to a web-server that can be accessed from a WebGL compatible browser on a mobile- or compute-based HMD.

We now describe the technical solutions underpinning our approach.

\subsection{{\sc WebVR} and the {\sc A-Frame} API}
We choose to create immersive VR experiences using the {\sc WebVR} {\sc A-Frame} API, developed by Mozilla's VR team.  Using {\sc html} statements, supported by Javascript functions, VR environments can be built quickly in a text editor.  The result can then viewed in any HMD that supports browser-based VR, provided {\sc WebGL}\footnote{\url{https://www.khronos.org/webgl/}} capabilities have been activated (usually selected in the browser preferences).

{\sc A-Frame} uses an entity-component architecture: one or more abstract modules can be attached to each element within a scene.  The components can alter the way an entity appears (size, colour, material), how it moves, or whether it reacts to other entities.   The entities and their components are described using {\sc html}  statements with a syntax modelled on the CSS (Cascading Style Sheets) language.\footnote{\url{https://www.w3.org/standards/webdesign/htmlcss}}  Standard {\sc A-Frame} entities exist for a number of geometrical primitives, such as spheres ({\tt <a-sphere>}), cylinders ({\tt <a-cylinder>}), and text ({\tt <a-text>}).

\subsection{{\sc A-Frame} geometrical primitives}
\label{sct:prim}
In most graphics libraries, the locations of objects can be shown using three main geometrical primitives: points, spheres or billboards.   A billboard comprises a low-resolution texture assigned to a rectangular polygon, often providing the option with the best visual quality.  Billboards are continuously reoriented so that they always point towards the camera, which requires computation on every refresh cycle.   The size and colour of all three primitives may be adjustable, along with the polygon resolution of the sphere.

Within an {\sc A-Frame} scene, points are represented as a single pixel, making them hard to see, and hence this option is not suitable.  Additionally, there is no textured billboard in the {\sc A-Frame} core -- although some third-party components have been developed.   Since there is more control over the appearance of spheres, particularly the polygon resolution and radius, it is preferable to use this as the primary primitive to display individual objects for the quickstart mode of {\tt allskyVR}.  As we discuss in Section \ref{sct:textures}, it is possible to use other primitives in the customisable version, but at the cost of some level of interactivity.

For finer control of each entity's appearance, we elect to use the more generic blank {\tt <a-entity>} form, and attach a geometry component, which assigns the sphere primitive type.   The polygon count for each of the sphere primitives is controlled by the {\tt segmentsWidth} and {\tt segmentsHeight}  parameters, allowing the user to select the trade-off between aesthetics (higher polygon count preferred) and performance (lower polygon count preferred).

Functionality in the {\tt allskyVR} distribution performs the conversion of each object's position, colour and scale factor to an {\sc A-Frame} entity declaration.

An example of the statements required to build a simple {\sc A-Frame} scene are shown in Figure \ref{fig:entities}.  Here, four red spheres are placed in front of the viewer in an environment surrounded by a dark blue spherical sky.   The statements defining the primitives are enclosed within an {\tt <a-scene>} hierarchy.

For use within an {\sc A-Frame} scene, $(\alpha, \delta, r)$ coordinates are converted to an $(x, y, z)$ Cartesian triple; the category index is used to place each object into a user-selectable hierarchy; and the scaling factor controls the relative size of the geometrical primitive used to show the object's location.

\subsection{Adding a celestial coordinate system}
\label{sct:SkyCube}
Mapping coordinates and colours to a collection of low-polygon count spheres is a first step towards an immersive all-sky experience. However, without a visible coordinate grid for reference, it is more difficult to orient oneself within the VR environment. 

Generating smooth line segments for constant lines of right ascension and declination can result in an unacceptably high polygon count.  Moreover, as there is no line width option, it is necessary to use cylinders if the line thickness needs to be controlled. These effects can have a significant impact on the level of interactivity and responsiveness of the display to head movements.     Instead of using line segments or cylinders, we use a pre-generated image of the coordinate system, which is wrapped around the viewer.

\begin{figure*}
\begin{center}
\includegraphics[width=1.00\linewidth,angle=0]{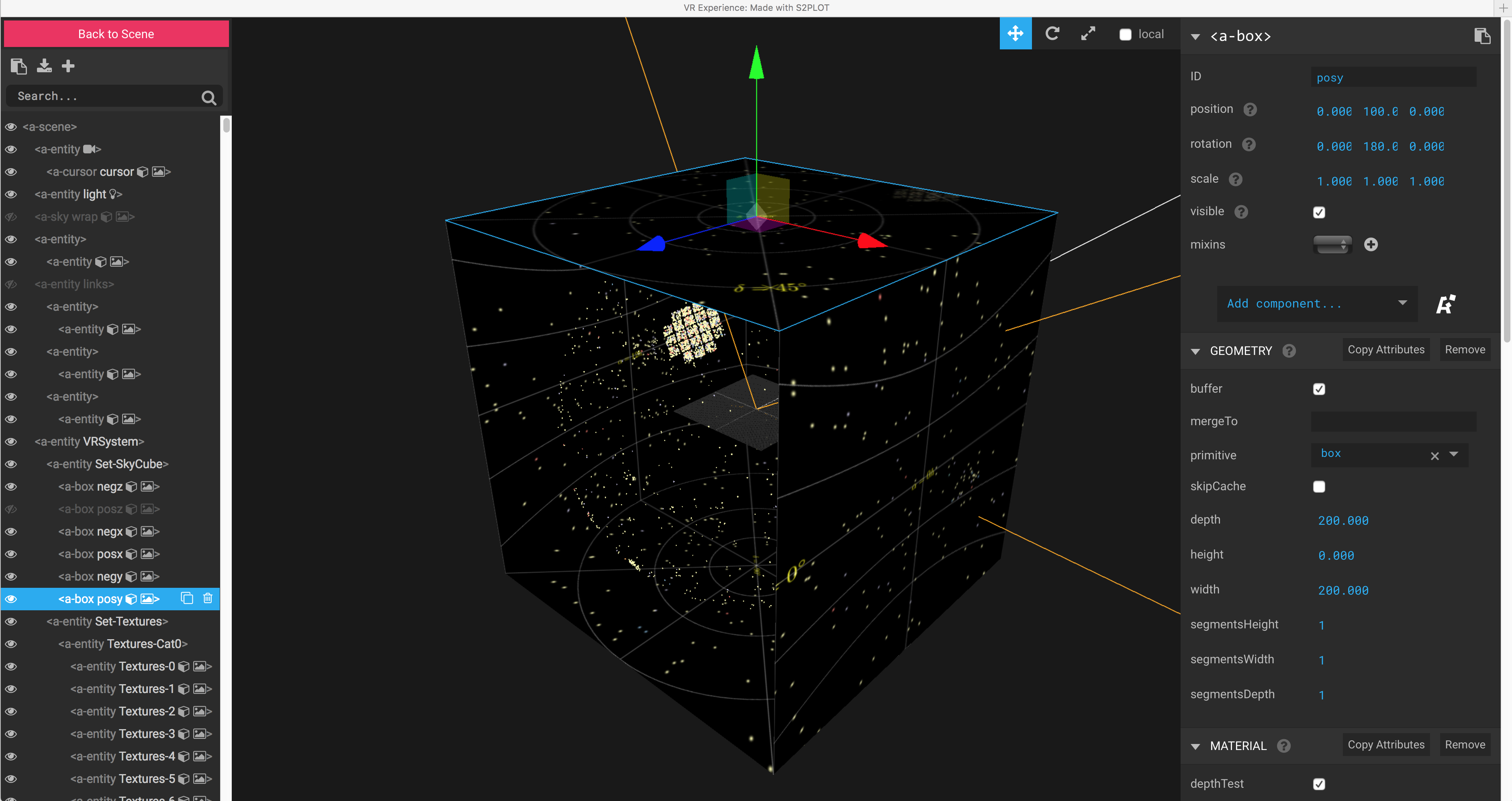}
\end{center}
\caption{The {\sc A-Frame} Inspector.  This view is accessed by pressing {\tt <Cntrl>-<Option>-I} when viewing outside of full-screen mode in a browser.  The textures forming the Sky Cube are visible: one of which ({\tt posy}) has been selected.  The {\tt posz} texture has been hidden by selecting the eye symbol from the hierarchy of entities on the the left-hand side.  This reveals the individual sphere entities inside the Sky Cube.  Attributes can be modified for each entity using the options on the right-hand side. The data set from the Kepler space mission is described in more detail in Appendix \ref{sct:example1}.}
\label{fig:inspector}
\end{figure*}

\subsubsection{Image-based coordinate grids}
The {\sc A-Frame} {\tt <a-sky>} entity maps an image with a 2:1 aspect ratio in equirectangular coordinates onto a sphere.
While there is minimal distortion along the celestial equator, the pixel density changes rapidly across the image.  Visually, this approach is unsatisfactory, particularly when looking towards the poles [e.g. as occurs with the YouTube spherical panorama movies -- see examples by \citet{Kent17} and \citet{Russell17}].   Moreover, the {\tt <a-sky>} entity is a tessellated sphere, which can increases the polygon count in the scene by a few thousand faces in order to achieve a sufficiently smooth surface.

To create a more complex, immersive all-sky experience for HMDs, without substantially increasing the polygon count, we use a technique introduced to computer graphics in the mid 1980s:  perspective projection onto the interior six faces of a cube.  We refer to this geometrical element as a Sky Cube.

The original approach, as described by \citet{Greene86}, was proposed as a computationally efficient method for environment mapping.  It allowed reflection effects and illumination effects from the environment surrounding a model to be included without the overheads of ray-tracing.

The primary limitation of the Sky Cube approach is that the viewer cannot navigate away from the origin: the projection is only correct when viewed from the same location at which it was generated.  However, this is the specific scenario we are trying to address.  As soon as any physical navigation other than head-orientation is used, or objects are presented at different radial values, the resulting mapping again distorts area, angle or separation.   

In principle, the Sky Cube should be placed infinitely far from the viewer.  In practice, a suitably large value for the cube dimensions is chosen, so as to be compatible with the limits of the graphics depth buffer.

\subsubsection{Building a Sky Cube with {\sc s2plot}}
\label{sct:textures}
We generate a Sky Cube by placing {\sc s2plot}'s virtual camera at the origin of the celestial sphere.  Using {\sc s2plot}'s dynamic callback system, the view-direction and camera up vector are modified on successive refresh cycles to produce six projections each with a 90 degree field of view.      The camera vectors are set using the {\tt ss2sc(...)} function, and the camera angle is set using the {\tt ss2sca(...)} function.   A one frame delay is required in the refresh cycle, as the {\sc tga} export function {\tt ss2wtga(...)} saves the previously-generated frame. 

To improve visual quality, texture-based axis labels and other annotations are generated using either \LaTeX\-style statements, {\tt ss2ltt(...)}, or with FreeType fonts, using the {\tt ss2ftt(...)} function.   The {\sc s2plot} environment variables listed in Appendix \ref{app:depend} must be set in order for these two texture types to be used.   

FreeType textures are preferred for text-only labels, as they provide a great deal of flexibility in the choice of font.   \LaTeX\ fonts allow the standard set of mathematical symbols and type-setting commands for superscript and subscript fonts, etc. to be used.

Due to the wide field-of-view of each face of the Sky Cube, we need to orient any textures towards the origin.  This is achieved by:
\begin{enumerate}
\item Querying the camera's up vector, ${\mathbf {u}}$, and unit view direction vector, $\mathbf{v}$, with {\tt ss2qc(...)};
\item Calculating a local right vector for a texture centred at $\mathbf{x}_i$ via the cross product: $\mathbf{r} = {\mathbf {u}} \times \mathbf{x}_i$; and
\item Calculating a new local up vector: $\mathbf{n} = {\mathbf {u}} \times \mathbf{x}_i$.
\end{enumerate}

The vectors ${\mathbf{r}}$ and ${\mathbf{n}}$ are converted to unit vectors and then used to determine the four vertices of a polygon onto which the relevant texture is mapped.  Some user adjustment of the label content, font and size may be required to ensure the best possible appearance of FreeType or \LaTeX\-based textures.

An advantage of using a Sky Cube is that additional geometrical primitives may be rendered into the six cube views.    Along with axis labels, this might include additional annotations, background imagery (not supported in the initial release of {\tt allskyVR}), or representations of object positions when there are too many items in a catalogue to display in real-time.   

A case where this might be relevant is when images of specific objects are to be ``baked'' into the Sky Cube views.  Here, for example, a package like Montage \citep{Berriman17} could be used to obtain a set of cut-out images which are then used as individual billboard texture maps within a customised {\sc s2plot} application.   However, as billboards are continuously oriented towards the camera, some care in interpretation is required, as the images may not always preserve their correct spatial relationships on the sky once baked into the Sky Cube.  

As this level of user-specific customisation will vary case-by-case, a sample workflow is included in Appendix \ref{sct:example1}.   A vanilla Sky Cube is included for the quickstart mode; alternative options can be downloaded from the {\tt allskyVR} web-site.

\subsection{Interaction}
\label{sct:menu}
It is usually impractical to type commands on a keyboard while immersed in a virtual environment. Not all commercial HMDs, however, are equipped with a controller for navigation. 

Smartphone-style VR systems, such as Google Cardboard,  provide limited selection-based interaction through a button integrated into the HMD housing.   The initial release of the Samsung Gear VR system provided a touch-pad on the side of the HMD, along with two buttons linked to menu actions.   

While {\sc A-Frame} provides a a tracking interface to HTC Vive and Oculus controllers through the {\tt vive-controls} and {\tt oculus-touch-controls} entities, we do not explore them further at this stage.   Instead, for convenience, we provide an easily customisable gazed-based menu system, which allows the user to make a hands-free selection of different parts of a scene.    Our solution remixes some elements of the {\sc A-Frame} $360^\circ$ image gallery example.\footnote{\url{https://aframe.io/examples/showcase/360-image-gallery/}} 

Physical attributes (e.g. mass, magnitude, morphology) may be supported by categorical labels, which lend themselves to a hierarchy that can be implemented by nesting {\sc A-Frame} entities.   
The menu system allows the visibility of named entities (and their children in the hierarchy) to be toggled.  The cursor (a green circle, whose default appearance can be modified in the {\sc html}  source) is drawn at the centre of the field of view.  
This achieved by using the Javascript {\tt addEventListener($\dots$)}, which then changes the visibility of the element through a call to {\tt setAttribute($\dots$)}.

As the viewer's head position changes, the cursor can be brought into alignment with a textured element for the menu.   Maintaining focus on this element causes a sub-menu to appear: selection from this new menu toggles visibility of the categories and the Sky Cube.   The default locations of the menu entities can be modified, as the textures may obscure important parts of the data set.  An optional format file can be specified at runtime, 
containing the category-based colours and the short text-only labels that will
appear in the menu.

Code in the {\tt allskyVR} distribution manages the creation of the {\sc A-Frame} hierarchy, textures for the menu system, and integrates the various assets into an output directory containing the {\sc html}  source file, images and Javascript.   

Additional customisation and interaction with the {\sc A-Frame} entities can be performed using the {\sc A-Frame} Inspector.  The Inspector is accessed by pressing {\tt <Cntrl>-<Option>-I}, but only when viewing outside of full-screen mode in a browser (Figure \ref{fig:inspector}).

\section{Concluding remarks}
\label{sct:discuss}
Standard astronomy software has not yet caught up with the availability of HMDs.  This makes it difficult for experimentation, or wider-scale early adoption, to occur.   Without an easy way to look at your all-sky data with a HMD, how can you objectively assess the level of insight or potential for discovery that could arise?

 For highly customisable astronomy visualisation solutions, it is probably necessary to learn how to use vendor specific Software Development Kits \cite[SDKs; e.g.][]{Schaaff15}, or create new solutions based on game engines such as {\sc Unity}\footnote{\url{https://unity3d.com}} \citep{Ferrand16} or {\sc Unreal Engine}.\footnote{\url{https://www.unrealengine.com}}

{\sc Unity} and {\sc Unreal Engine} offer a powerful collection of primitives, management of dynamic timeline-based events, and support for HMDs and their corresponding interaction/motion detection solutions.   However, they require some familiarity with concepts and approaches from computer generated imagery (CGI) modelling and animation.  

The use of a hardware-specific SDK comes with a steep learning curve, and can lock a developer to a single vendor.  While this might be appropriate for commercial products (VR games and other entertainment experiences), it is less suitable for a generic solution that can be adopted by researchers.  Minor changes in any SDK can have a significant impact on development.

\begin{figure}
\begin{center}
\includegraphics[width=1.00\linewidth,angle=0]{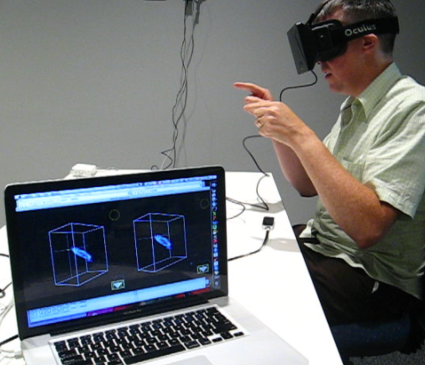}
\end{center}
\caption{An early experiment with the Oculus Rift DK1 and a Leap Motion controller through a custom {\sc s2plot} application. Note the image pair rendered on the laptop (lower-left), but without any attempt to correct the lens-based distortions.}
\label{fig:oculusDK1}
\end{figure}

As an example, we performed some simple experimentation with an Oculus Rift DK1 headset in 2015.   Using  {\sc s2plot}, and an incomplete integration with the Oculus API, it was possible to generate a usable virtual experience.  In Figure \ref{fig:oculusDK1}, the image pair displayed in the HMD can be seen on the laptop screen in the lower left of this image.
and the faint (yellow) circle shows the finger position detected by a Leap Motion\footnote{\url{https://www.leapmotion.com}} controller.   
No attempt was made to correct the significant chromatic distortions introduced by the DK1 lenses.  Unfortunately, support for macOS and Linux operating systems was paused by Oculus soon afterwards due to concerns about the minimum level of graphics performance required to power the Rift.\footnote{See, for example, \url{https://www3.oculus.com/en-us/blog/powering-the-rift/}}  Consequently, we abandoned this approach.

The {\tt allskyVR} approach provides a pathway to early adoption of immersive VR for visual exploration and publication.
It performs a limited number of tasks related to conversion from a simple input data format to a VR experience, where the only mode of navigation is based on view-direction.   When an {\sc A-Frame} environment is viewed with a compatible HMD, it is possible to explore spatial relationships between objects from a user-centered spherical coordinate system.

We recognise that every astronomer using data in spherical coordinate systems will have different requirements.   In our solution, we provide two pathway to adoption of HMDs: a quickstart solution, where conversion of data values to the  relevant {\sc html} statements can be achieved with any number of approaches, and a fully-customisable solution for those who can see the benefit of the {\sc s2plot} programming library to provide more control over the visual outcome.   As we have favoured functionality over efficiency in {\tt allskyVR},  we welcome suggestions to improve the run-time performance if this becomes an issue for some users.  Our own testing suggests that interactivity is maintained for datasets comprising a few 1000 objects rendered as spheres, however, this is device and screen-resolution dependent.

In order to provide astronomers with opportunities to explore their all-sky data with a HMD, a simplified, vendor-agnostic process is required.   {\sc A-Frame} shows one simple way to present all-sky catalogues within a virtual environment, which can be viewed with variety of HMDs.  We look forward to seeing how the astronomy-focused virtual reality evolves, and where it finds its niche within the complementary phases of visual exploration and publication.

\bibliographystyle{pasa-mnras}
\bibliography{immersive}

\appendix
\section{{The \sc s2plot} library and dependencies}
\label{app:depend}
{\sc s2plot} \citep{Barnes06} is an advanced, three-dimensional graphics library built as a layer on top of OpenGL.   This open source library is used most effectively within C/C++ programs. Its key features include a rich application programming interface and simple support for a variety of standard and advanced displays including: side-by-side, frame sequential, and interlaced stereoscopic modes; and full/truncated fish-eye projections suitable for digital domes. 

{\sc s2plot} provides a variety of geometrical primitives ranging from points, lines, and spheres to isosurfaces and volume renderings, most of which can be created or displayed with a single function call.  Labels or text annotations can be created using \LaTeX\ and FreeType fonts.  Geometrical primitives can either be  created as static objects (these are created once, and are always displayed) or as dynamic objects (these  are re-generated at each screen refresh). 

Interactive inspection of 3D data sets is achieved with mouse and keyboard controls, with default behaviour for many key-presses (auto-spin the camera, zoom in/out) and user customisation via a {\em callback} system.  

To access the full functionality presented in this paper requires a working installation of the {\sc s2plot} distribution (version 3.4 or higher), available from:
\begin{center}
{\tt https://github.com/mivp/s2plot} 
\end{center}
along with the following additional items:
\begin{itemize}
\item The ImageMagick\textregistered  tools\footnote{\url{http://www.imagemagick.org}}: the {\sc convert} utility is used throughout to make conversions between image formats supported by {\sc s2plot} ({\sc tga}) and \LaTeX\ ({\sc png}); and
\item The {\tt allskyVR} bundle: comprising the C-language source code, header files, build scripts, and a directory containing pre-generated {\sc A-Frame} assets. 
\end{itemize}

The {\sc s2plot} environment variables {\tt S2PLOT\_IMPATH}, {\tt S2PLOT\_LATEXBIN}, {\tt S2PLOT\_DVIPNGBIN} must all be set as described in the {\tt ENVIRONMENT.TXT} file included in the {\sc s2plot} distribution.    If FreeType textures are to be used for axis labels and other annotations (see Section \ref{sct:textures}), the FreeType libraries needs to be installed and the {\tt S2FREETYPE} environment variable  set to {\tt yes}.   It may be necessary to modify the {\tt \_DEFAULTFONT} variables defined in {\tt allskyVR.h} to better reflect a given system configuration or to manage a specific use case. 

For VR export to work correctly, {\tt S2PLOT\_WIDTH} and {\tt S2PLOT\_HEIGHT} must be set to the same value.   It is recommended that the largest possible square window is used for the best graphics quality.

\section{Star systems with confirmed exoplanets}
\label{sct:example1}
In this Appenix, we provide a step-by-step guide to creating an immersive VR environment using {\tt allskyVR}.   As an example, we populate the celestial sphere with the locations of confirmed exoplanet systems.

The first step is to gather data, and convert it to an appropriate format.   We access the NASA Expolanet Archive\footnote{\url{https://exoplanetarchive.ipac.caltech.edu}}, and view the table of confirmed planets.  Selecting the $\alpha$, $\delta$, and number of planets ($N_{\rm p}$) columns, the data set is downloaded and saved in comma-separated variable ({\sc csv}) format.    Some editing is required to remove duplicate items and the columns of sexagesimal-formatted coordinates.  A new data column is created to hold the  $r_z$ coordinates, with all values set to 1.    

The {\em Kepler} spacecraft's primary mission footprint, which contributes a substantial number of exoplanets to the data set, requires some additional attention.  A further column is created to contain the relative scaling sizes, $S_{\rm g}$, for the geometrical primitives.    Exoplanet systems with celestial coordinates in the range  $18 {\rm h} \leq \alpha \leq 22 {\rm h}$ and $+30^\circ \leq \delta \leq +60^\circ$ have a scale value of 1, while other systems have a scale value of 5.  This will mean that isolated systems are more easily seen, while limiting the over-crowding in the Kepler field.     For other data sets, additional fine-tuning may be required to produce the most effective visualisation.

This modified data set is saved as a {\sc csv}-format file ({\tt exoplanet.csv}), with the column order: $\alpha$ (decimal degrees), $\delta$ (decimal degrees), $r_z$, $N_{\rm p}$, $S_{\rm g}$.  

User control of colours and tags is provided through a text file, {\tt format.txt}, requiring one line in the file per category.  The format is:
\begin{verbatim}
CAT=R,G,B,Label
\end{verbatim} 
where the red, green, and blue colour components, [{\tt R,G,B}], are integer values in the range
[0..255], and {\tt Label} is a short text-only label to appear in the {\sc A-Frame} menu.  
It is necessary to avoid spaces and some symbols in the label, such as {\tt \$} and ${\tt \_}$, which have particular meanings in \LaTeX\ formatting.

Note that the category labels are only used in the fully customisable mode (Section \ref{sct:textures}), where the relevant textures are generated on demand.  For the quickstart mode, default textures are provided from an asset directory.  These can be replaced by the user as required.

After setting the {\tt S2PLOT\_WIDTH} and {\tt S2PLOT\_HEIGHT} environment variables to 800 (pixels),
we launch the {\sc s2plot} application with the following command-line arguments:
\begin{verbatim}
templateSpherical -i exoplanet.csv 
        -f format.txt -o exosys 
\end{verbatim}

On pressing {\tt <shift>-v}, the virtual reality experience is generated, with all assets moved to the directory {\tt VR-exosys}.  The exoplanetary system data set is now ready to be deployed on a relevant web-server for viewing with a compatible head-mounted display.  The two panels in
Figure \ref{fig:VRexo} show screenshots captured from a Samsung Galaxy S7 Edge 
mobile device.  The expanded menu, including the category labels, is visible in the lower panel.

\subsection{Modifying the Sky Cube textures}
Suppose the input catalogue comprised a few thousand additional planetary candidates, thus reducing the responsiveness of the immersive experience,  and it was sufficient to only interact with a sub-set of categories.  The following workflow would allow a user-controlled portion of objects to have their positions ``baked'' into the Sky Cube texture.   
\begin{enumerate}
\item Create two input data files, one which contains only the data items that will be ``baked'' into the Sky Cube (exoplanet-bake.csv), and one containing only the data items for interactive exploration (exoplanet-interact.csv).   Visibility of the latter will be controllable from the HMD using the gaze-based menu system, so use of relevant categories to further subset the data is encouraged.   
\item Execute the {\sc s2plot} application using the smaller of the two data sets, and complete the export step:
\begin{verbatim}
templateSpherical -i exoplanet-interact.csv 
        -f format.txt -o exosys 
\end{verbatim}
This will create the relevant {\sc html}, Javascript and {\sc a-frame} for interactive exploration. 
\item Make a back-up (in another location) of the six Sky Cube textures: {\tt neg?.png} and {\tt pos?.png}.
\item Execute the {\sc s2plot} application using the larger of the two data sets, choosing a different export directory name, to avoid over-writing the first export:
\begin{verbatim}
templateSpherical -i exoplanet-bake.csv 
        -f format.txt -o exosys-bake
\end{verbatim}
This will create the Sky Cube textures with the additional data set items included.
\item Copy the new set of six Sky Cube textures from the {\tt VR-exosys-bake} directory to the {\tt VR-exosys} directory, replacing the original textures.
\end{enumerate}
The result is an immersive environment with a more detailed Sky Cube, which can still
have its visibility toggled through the menu system.

\end{document}